\pdfoutput=1
\documentclass[aps,pra,twocolumn,showpacs%,superscriptaddress
]{revtex4-1}
\usepackage{amssymb}

\usepackage{graphicx,subfigure}

\usepackage[sort&compress]{natbib}

\begin{document}

\title{Mode pairs in $\mathcal{PT}$-symmetric multimode waveguides}
\date{\today}
\author{Changming Huang}%$^1$
\author{Fangwei Ye}
\email{fangweiye@sjtu.edu.cn}
\author{Xianfeng Chen}

\affiliation{State Key Laboratory of Advanced Optical Communication Systems and Networks, Department of Physics and Astronomy, Shanghai Jiao Tong University, Shanghai 200240, China}

\date{\today}

\begin{abstract}
We study mode properties in multimode optical waveguides with parity-time($\mathcal{PT}$) symmetry. We find that two guiding modes with successive orders 2\emph{m}-1 and 2\emph{m} form a mode pair in the sense that the two components of the pair evolve into the same mode when the loss and gain coefficient increases to some critical values, and they experience $\mathcal{PT}$ symmetry breaking simultaneously. For waveguides that in their conservative limit support an odd number of guiding modes, a new mode with a proper order emerges upon the increase of the gain/loss level, so that it pairs with the already existing highest-order mode and then break their $\mathcal{PT}$ symmetry simultaneously. Depending on the specific realizations of $\mathcal{PT}$-symmetric potentials, higher-order mode pair may experience symmetry breaking earlier or later than the lower-order mode pairs do.
\end{abstract}

\maketitle

%\preprint{}




%\thanks{}



%\keywords{}

\section{INTRODUCTION}
One of the postulates of quantum mechanics is that every physical observable corresponds to a Hermitian operator so that the eigenvalues are guaranteed to be all pure real. However, Bender and co-workers revealed that a non-Hermitian Hamiltonian respecting the so-called parity-time($\mathcal{PT}$) symmetry can still exhibit an entirely real spectra\cite{Bender1,Bender2,Bender3}. By definition, a Hamiltonian is said to be $\mathcal{PT}$-symmetric if it shares a common set of eigenfunctions with the $\mathcal{PT}$ operator. The parity operator $\mathcal{P}$, responsible for spatial reflection, is defined through the operations $P \rightarrow -P, x \rightarrow -x$, while the time reversal operator $\mathcal{T}$ leads to $P \rightarrow -P, x \rightarrow x$ and to complex conjugate $i \rightarrow -i$. Given the fact $\mathcal{TH}=P^2/2+V^*(x)$, a necessary condition for the Hamiltonian to be $\mathcal{PT}$-symmetric is that the potential function $V(x)$ should satisfy the condition $V^{*}(-x)=V(x)$. However, the latter is only a necessary condition for $\mathcal{PT}$ symmetry, because the transition to a complex spectrum, which is called the $\mathcal{PT}$-symmetry breaking, appears upon the increase of the strength of the imaginary part of the potential $V(x)$.

Optical structures are suggested to be a powerful platform for the implementation of $\mathcal{PT}$ physics \cite{JPA2005,DNPRL2008,Visualization,LonghiPRL,new4}. Spontaneous $\mathcal{PT}$ symmetry breaking has been experimentally observed in passive\cite{Exp1} and active \cite{Exp2} optical waveguide couplers. Since then various $\mathcal{PT}$-symmetric structures were studied, including nonlinear couplers\cite{nlcoupler1,nlcoupler2,nlcoupler3,nlcoupler4,nlcoupler5,nlcoupler6}, periodic\cite{lattice1,lattice2,lattice3,lattice4,lattice5,lattice6,LonghiPRL} or truncated\cite{truncated1,truncated2,truncated3} and defective lattices\cite{defected1,defected2}, pseudo-potentials with $\mathcal{PT}$-symmetric \cite{nonlinearpt1,nonlinearpt2} or inhomogeneous nonlinear terms \cite{YVK2014} , as well as mixed linear and nonlinear lattices\cite{mixed1,mixed2,new3}.

In this paper, with reference to waveguiding structures that support a large number of localized modes, we put forward a systematic study on the $\mathcal{PT}$-symmetry properties of higher-order modes. We find that, modes with successive order 2\emph{m} and 2\emph{m}-1 form a \emph{mode pair}(\emph{m} is a positive integer), as the two components of the pair evolve into the same mode when the gain/loss level increases to some critical values, and they simultaneously break their symmetry. Interestingly, in the case when the waveguides in their conservative limit accommodate an odd number of modes, the increase of the gain/loss level creates a new mode whose order is larger by one than the already-existing highest-order mode, with which the new mode pairs and they experience $\mathcal{PT}$ symmetry breaking simultaneously. We also find that, depending on the specific realizations of $\mathcal{PT}$-symmetric potentials, the critical value of the gain and loss coefficient beyond which the symmetry of the higher-order mode pair breaks could be larger or smaller than those of the lower-order mode pairs. It should be noted that the simultaneous symmetry breakup of the fundamental and dipole modes has been reported in Refs.\cite{new1,new2}, however, generic properties of higher-order modes have not been systematically studied, and the latter is the aim of this paper.
\begin{figure*}[htbp]
\centering
%\subfigure{\label{fig:fig1a}\includegraphics[width=0.24\textwidth]{fig1a.png}}\hspace{-0.01\textwidth}
%\subfigure{\label{fig:fig1b}\includegraphics[width=0.24\textwidth]{fig1b.png}}\vspace{-0.025\textwidth}\\
%\subfigure{\label{fig:fig1c}\includegraphics[width=0.24\textwidth]{fig1c.png}}\hspace{-0.01\textwidth}
%\subfigure{\label{fig:fig1d}\includegraphics[width=0.24\textwidth]{fig1d.png}}\vspace{-0.025\textwidth}\\
%\subfigure{\label{fig:fig1e}\includegraphics[width=0.24\textwidth]{fig1e.png}}\hspace{-0.01\textwidth}
%\subfigure{\label{fig:fig1f}\includegraphics[width=0.24\textwidth]{fig1f.png}}
\subfigure{\label{fig:fig1a}\includegraphics[width=13cm]{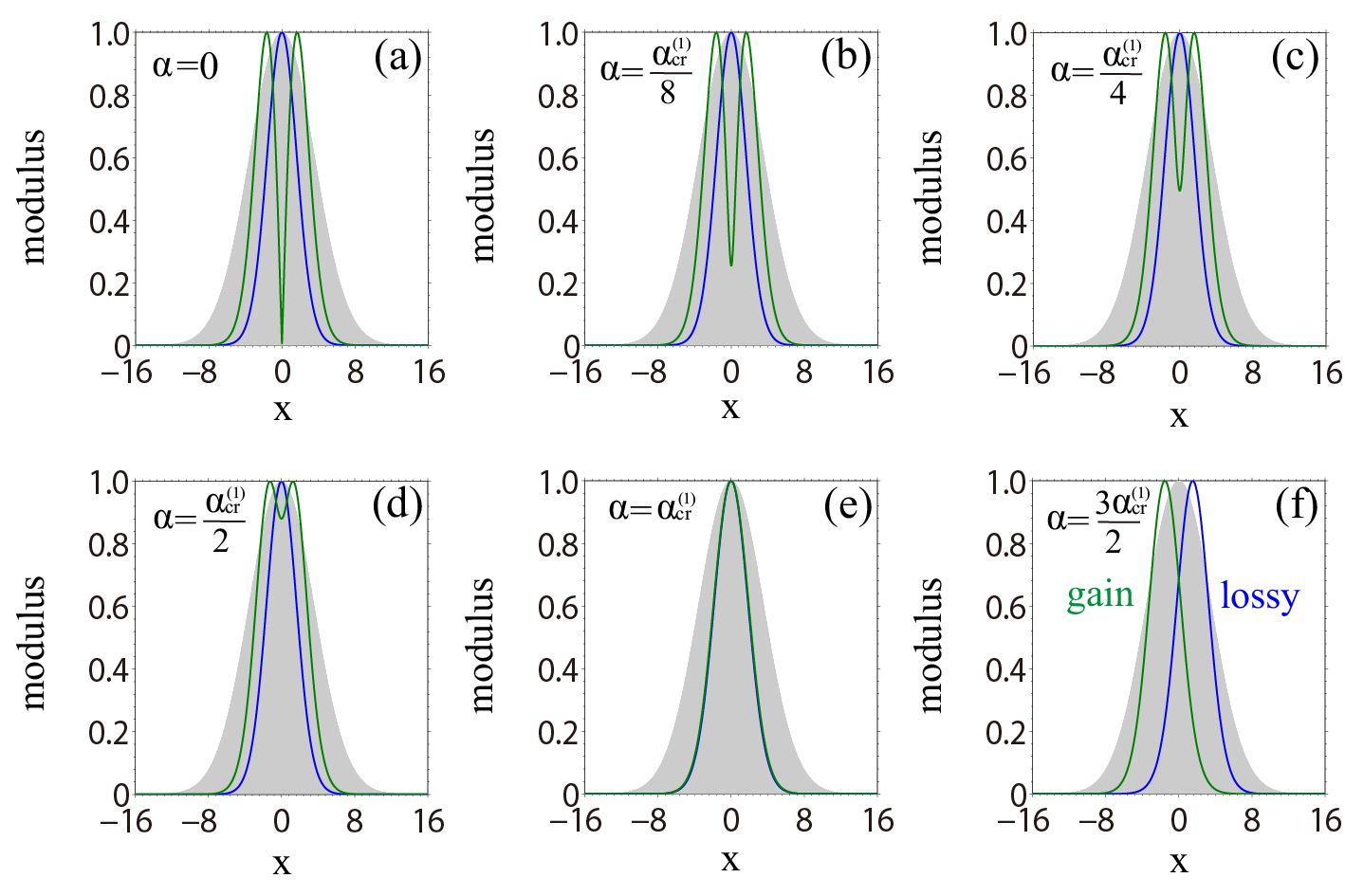}}\hspace{-0.01\textwidth}
\caption{\label{fig:Fig1} (Color online) Mode profiles for fundamental and dipole modes at different values of $\alpha$. The gray region stands for the landscape of the gaussian waveguide. $p=2,d=5$.}
\end{figure*}

\section{MODEL}
Let us consider the propagation of a laser beam in a multimode waveguide that can be described by a Schr\"{o}dinger-like equation for the dimensionless field amplitude $q$,
%\section{model}

\begin{equation}\label{eq:eq1}
  \centering
    i\frac{\partial q}{\partial z}=-\frac{1}{2}\frac{\partial^2 q}{\partial x^2}- V(x)q.
\end{equation}

Here $x$ and $z$ are normalized transverse and longitudinal coordinates, respectively. The complex function $V(x)$ describes the waveguide profile, whose real part represents the landscape of the refractive index, while imaginary part represents the gain and loss modulations.  While other types of waveguide profiles have also been checked in our study, for demonstration purpose, we assume in the following $V(x)=p \exp(-\frac{x^2}{d^2})(1+i\alpha x)$, namely, a Gaussian waveguide with a balanced gain and loss built into the waveguiding region. Parameter $p$ and $d$ characterize the amplitude and width of the waveguide respectively, and $\alpha$ is the gain and loss coefficient. The eigenmodes of the complex waveguides can be found numerically by looking for the solution of Eq.(1) in the form of $q(x,z)=w(x)\exp(ibz)$, where $w=w_r+iw_i$ are mode wavefunctions that are generally complex functions(in the no gain and no loss limit, the wavefunction can be chosen to be pure real), and $b=b_r+ib_i$ are the mode propagation constants.  With a proper setting of $p$ and $d$, the waveguide in its conservative limit supports multiple modes, with the first order mode being nodeless in profile and the $\emph{N}^{\text{th}}$-order mode featuring \emph{N}-1 nodes. Without loss of generality, we following set $p=2$ and $d=5$, and the corresponding waveguide in its vanishing gain and loss limit supports six modes. We then gradually increase loss and gain coefficient, and watch the evolutions of these modes. The results are presented below.

\section{MODE PAIRS}
Figure 1 shows the evolutions of fundamental and dipole modes. When $\alpha=0$, fundamental mode is bell-shaped and dipole mode features two sharp peaks with a node in between them(Fig.1(a)). However, with the increase of $\alpha$, the valley between the two peaks starts climbing steadily and eventually at $\alpha=\alpha_{\text{cr}}^{(1)}$, the valley vanishes and the dipole mode becomes bell-shaped, taking exactly the same shape as fundamental mode[Fig.~1(e)]! If $\alpha$ increases further, two asymmetric modes occur, with one mostly residing at the lossy region (labeled "lossy" in Fig.~1(f)), and the other mostly at the gain region (labeled "gain" in Fig.~1(f)).

The variations of propagation constants accompanying with such mode reshaping is shown in Fig.~2. As expected, the propagation constants of the fundamental and
 dipole modes remain pure real until the loss/gain level exceeds some critical value. However, with the increase of $\alpha$ from zero, the propagation constants of two
 modes approach each other, and finally, they merge into one at $\alpha=\alpha_{\text{cr}}^{(1)}$. This is the point that dipole evolves into bell-shaped and attains
 the same shape as fundamental mode. Beyond this point, a pair of complex conjugate $b$ emerges (Fig.~2), corresponding to the gain and lossy mode pair shown in Fig.~1(f).
\begin{figure}[htbp]
\centering
\subfigure{\label{fig:fig2a}\includegraphics[width=8cm]{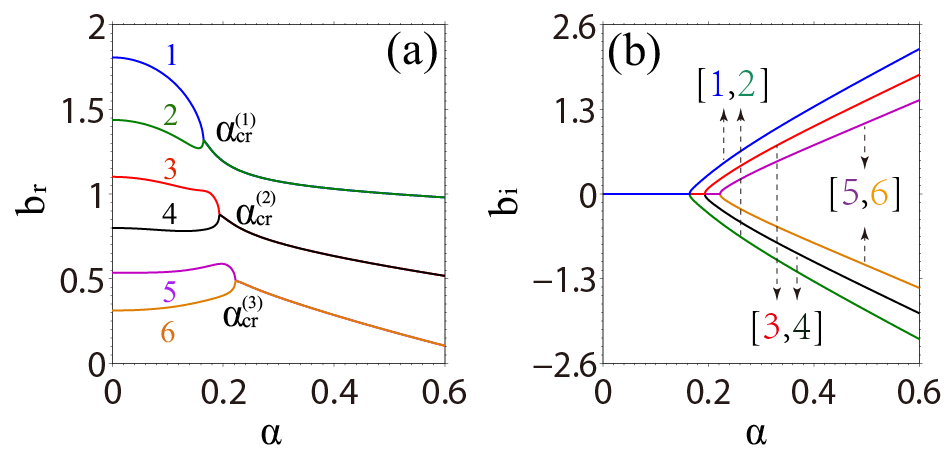}}\hspace{-0.01\textwidth}
\caption{\label{fig:Fig2} (Color online) Dependence of real(a) and imaginary(b) part of the propagation constants of the first six modes on $\alpha$.  $p=2,d=5$.}
\end{figure}

The approaching of the two components of the mode pairs can be understood analytically if one performs a perturbation analysis on Eq.~(1). The substitution of  $q(x,z)=w(x)\exp(ibz)$ into the equation yields the following equation for the stationary function, $w(x)$,
\begin{equation}
-bw+\frac{1}{2}w^{\prime \prime }+(V_{r}+i\beta V_{r}^{\prime })w=0,
\label{w}
\end{equation}%
where the prime stands for $d/dx$,\ $\beta \equiv -\alpha d^{2}/2$ is a
constant, and $V_{r}(x)$ is the real part of the potential (its particular
form is not important; it is essential that the imaginary part of the
potential is proportional to the derivative of the real part). \ For small $\alpha$ (thus also small $%
\beta $), solution is looked for perturbatively, as%
\begin{equation}
w(x)=w_{0}(x)+i\beta w_{1}(x),  \label{www}
\end{equation}%
where functions $w_{0}$ and $w_{1}$ are real, the zero-order function $w_{0}$
satisfies the usual linear Schr\"{o}dinger equation:%
\begin{equation}
-bw_{0}+\frac{1}{2}w_{0}^{\prime \prime }+V_{r}(x)w_{0}=0,  \label{w0}
\end{equation}%
and the first-order function $w_{1}$ satisfied the following inhomogeneous equation:%
\begin{equation}
-bw_{1}+\frac{1}{2}w_{1}^{\prime \prime }+V_{r}(x)w_{1}=-V_{r}^{\prime
}(x)w_{0}.  \label{w1}
\end{equation}

Now, applying $d/dx$ to Eq. (\ref{w0}), one obtains,%
\begin{equation}
-bw_{0}^{\prime }+\frac{1}{2}\left( w_{0}^{\prime }\right) ^{\prime \prime
}+V_{r}(x)w_{0}^{\prime }=-V_{r}^{\prime }(x)w_{0}.  \label{w0'}
\end{equation}%
Comparing Eqs. (\ref{w0'}) and (\ref{w1}) makes it obvious that
\begin{equation}
w_{1}(x)\varpropto w_{0}^{\prime }.  \label{10}
\end{equation}

\begin{figure}[htbp]
\centering
\subfigure{\label{fig:fig5a}\includegraphics[width=8cm]{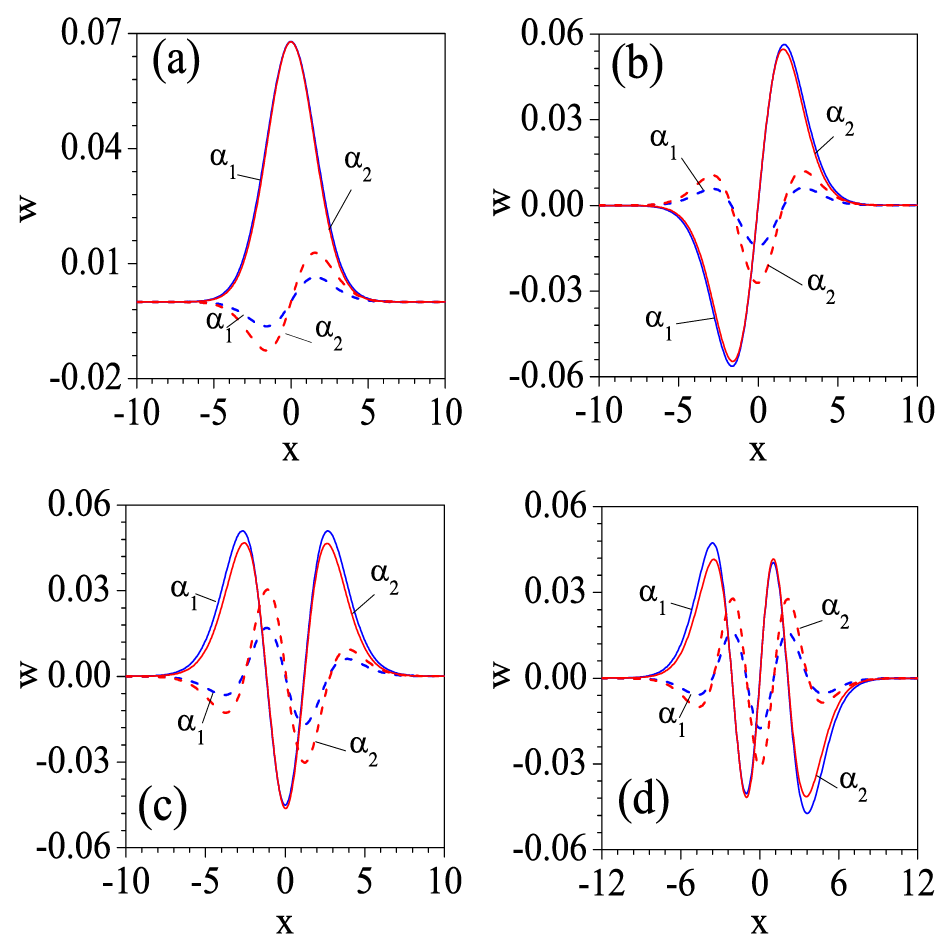}}\hspace{-0.01\textwidth}
\caption{\label{fig:Fig1} (Color online) The real (solid lines) and imaginary (dashed lines) parts of the  wavefunctions for the first (a, b) and the second (c, d) mode pairs,
 at $\alpha_1=0.02$(green lines) and $\alpha_2=0.04$ (red lines). $p=2, d=5$.}
\end{figure}

Thus, while fundamental and dipole mode differs significantly in their shapes
(actually they are orthogonal each other), gradually increasing $\alpha$ from zero introduces imaginary part into the mode wavefunctions. Importantly,
as expressions (3) and (7) show, the imaginary parts of the $\mathcal{PT}$-symmetric mode wavefunctions are proportional to the first derivative of their real parts,
with $\alpha$ being the proportional factor. Thus, the fundamental modes that are initially symmetric now gain imaginary parts that are antisymmetric, which
are exactly like the dipole modes. Similarly, the dipole modes that are initially antisymmetric now gain imaginary parts that are symmetric like a
fundamental modes. This prediction is well collaborated by numerically results (see Fig.~3). In another words, the increasing imaginary part of potential leads to
the growing weight of the second mode in the field of the mode that was initially "first", and the growing weight of the first mode in the field of the mode that
was initially "second". Such modes are not orthogonal any more and actually they start to approach each other. Thus, it is not surprising that, when the imaginary part
grows up to some critical value, fundamental and dipole modes take the same profiles and coalesce.

The mode approaching and eventually simultaneously symmetry breaking for dipole and fundamental modes are also observed for tripoles and quadrupoles, for $5^{\text{th}}$- and $6^{\text{th}}$-order modes, and so on. These are multi-humped modes with several valleys between the humps. Interestingly, the increase of $\alpha$ continuously
lifts up those valleys and weakens the amplitude modulations [Fig.~4(a,b,c)]. As a result, when $\alpha$ increases to $\alpha_{\text{cr}}^{(2)}$, tripoles and quadrupoles
evolve into the same bell-shape [Fig.~4(d)], and attain the same propagation constant (Fig.~2). The $5^{\text{th}}$- and $6^{\text{th}}$-order modes exhibit a similar
scenario (Fig.~2). After also performing the analysis on other types of $\mathcal{PT}$-symmetric multimode waveguides, we arrive at a conclusion that,
in multimode optical waveguides, guided modes with order 2\emph{m} and 2\emph{m}-1 form a \emph{mode pair} in the sense that the two components of the pair evolve into
the same mode at $\alpha_{\text{cr}}^{(m)}$ and they simultaneously undergo $\mathcal{PT}$ symmetry breaking beyond that point.

\begin{figure}[htbp]
\centering
\subfigure{\label{fig:fig3a}\includegraphics[width=8cm]{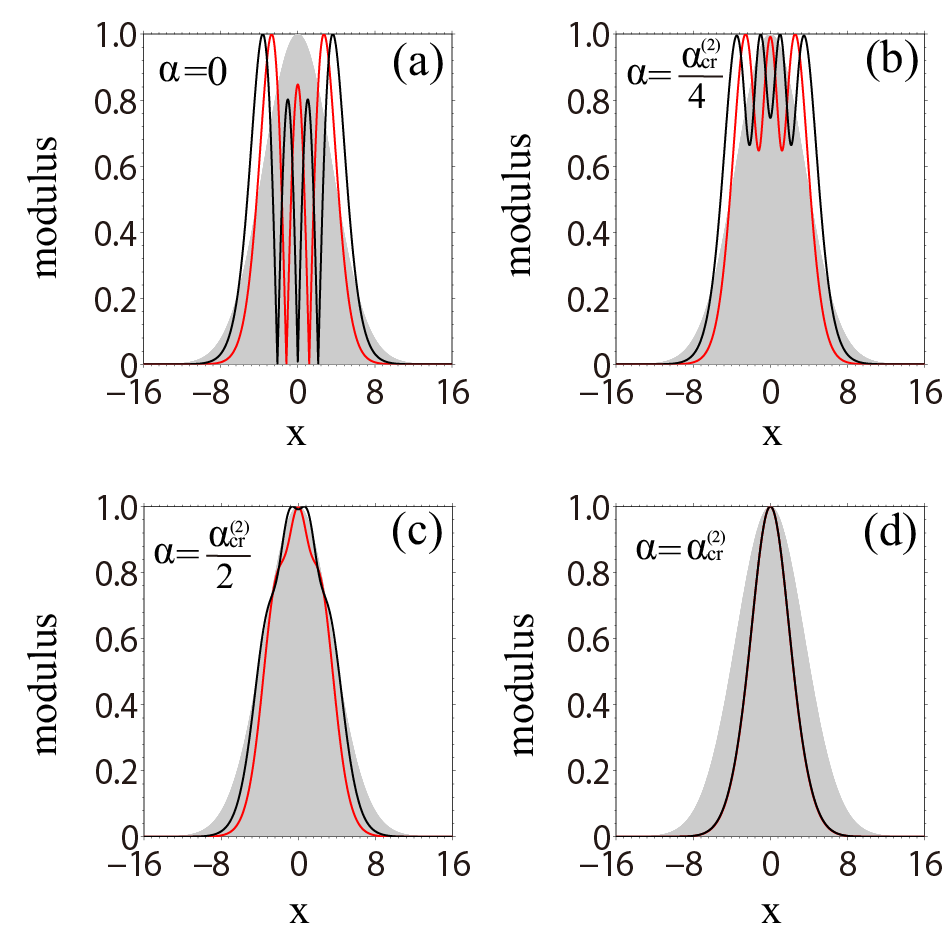}}\hspace{-0.01\textwidth}

\caption{\label{fig:Fig1} (Color online)  Mode profiles for tripole and quadrupole modes at different values of $\alpha$.  $p=2,d=5$.}
\end{figure}

\begin{figure}[htbp]
\centering
\subfigure{\label{fig:fig4a}\includegraphics[width=8cm]{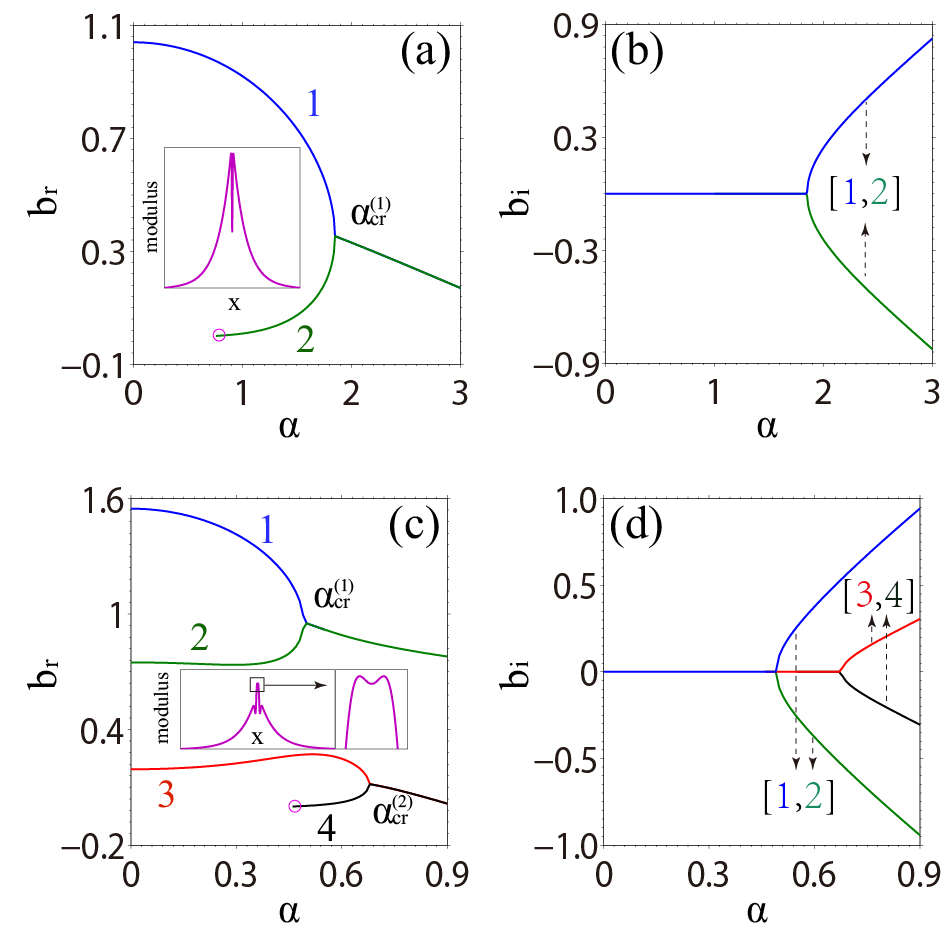}}\hspace{-0.01\textwidth}
\caption{\label{fig:Fig1} (Color online) Dependence of mode propagation constants on $\alpha$ for waveguides with $p=2,d=0.8$ (a,b) and $p=2,d=2$ (c,d).
 Insets in (a) and (c) show the mode profiles corresponding to the circles in the spectrum.}
\end{figure}

\section{MODE PAIRS IN WAVEGUIDES SUPPORTING ODD-NUMBERED MODES}

  The proposed concept of \emph{mode pair} naturally leads to the following question: what happens if a waveguide initially supports an odd number of guided modes,
   so that its highest-order mode, say, with the order of 2\emph{m}-1, does not have a chance to form a pair? Our thorough studies reveal that, the increase
   of $\alpha$ results in the formation of the 2\emph{m}$^{\text{th}}$-order mode, which pairs with the already existing (2\emph{m}-1)$^{\text{th}}$-order mode,
    and eventually, like other canonical mode pairs, this new pair enters the symmetry broken phase. Figure 5 illustrates the spectrum when the waveguide in its
    conservative limit accommodates only one (Fig.~5(a)) and three (Fig.~5(c))guided modes respectively. The figure shows that, at the early stage of the increasing
     $\alpha$, the fundamental (Fig.~5(a,b)) or tripole (Fig.~5(c,d)) evolves by itself without forming a pair with other modes. Interestingly, however, when
     $\alpha$ increases to some value, a new guided mode featuring two (Fig.~5(a)) or four(Fig.~5(c)) humps appears, which is recognized as a dipole or
     quadrupole mode. The new mode pairs with the already existing fundamental or tripole mode, and they experience a simultaneous $\mathcal{PT}$ symmetry breaking
     with the further increase of $\alpha$.

  \begin{figure}[htbp]
\centering
\subfigure{\label{fig:fig5a}\includegraphics[width=8cm]{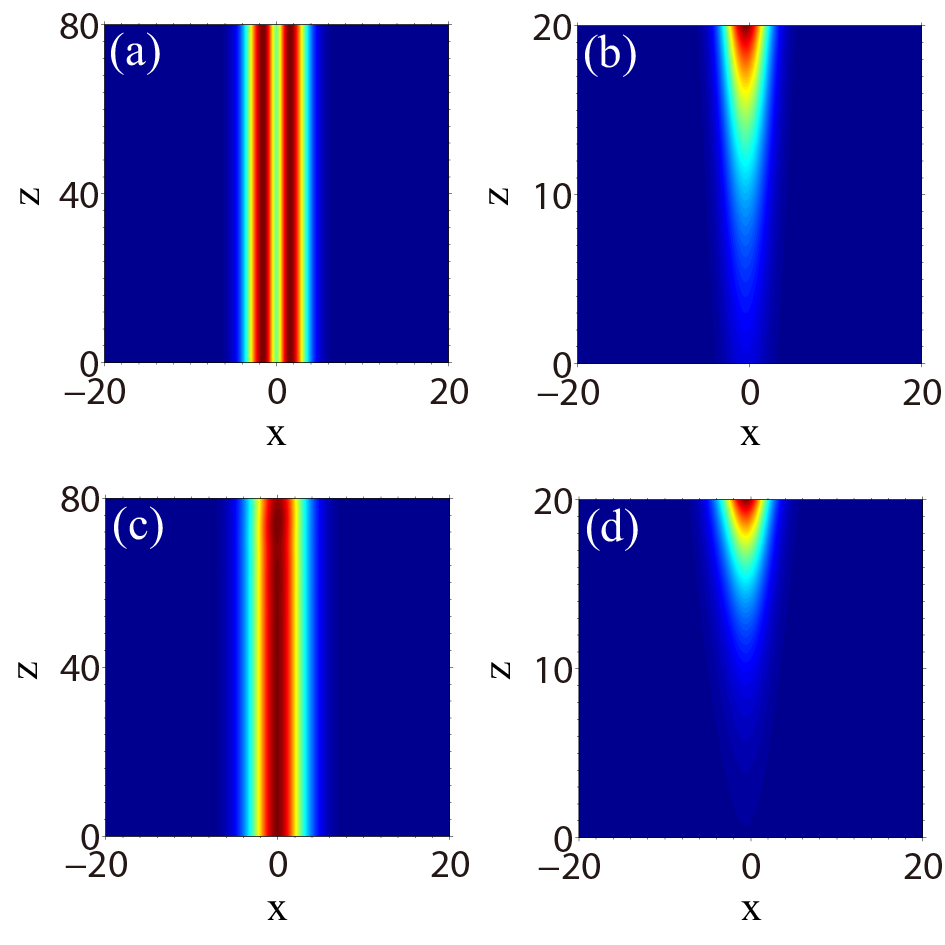}}\hspace{-0.01\textwidth}
\caption{\label{fig:Fig1} (Color online) Propagation simulation of the dipole mode at $\alpha=0.041$( $<$ $\alpha_{\text{cr}}^{(1)})$(a),
and of the mode bifurcating from fundamental/dipole mode pair at $\alpha=0.175(>\alpha_{\text{cr}}^{(1)})$(b). Propagation simulation of
the quadrupole mode at $\alpha=0.175(<\alpha_{\text{cr}}^{(2)})$(c), and of the mode bifurcating from tripole/quadrupole mode pair at
$\alpha=0.22(>\alpha_{\text{cr}}^{(2)})$(d). $p=2,d=5$.}
\end{figure}

\begin{figure}[htbp]
\centering
\subfigure{\label{fig:fig7a}\includegraphics[width=8cm]{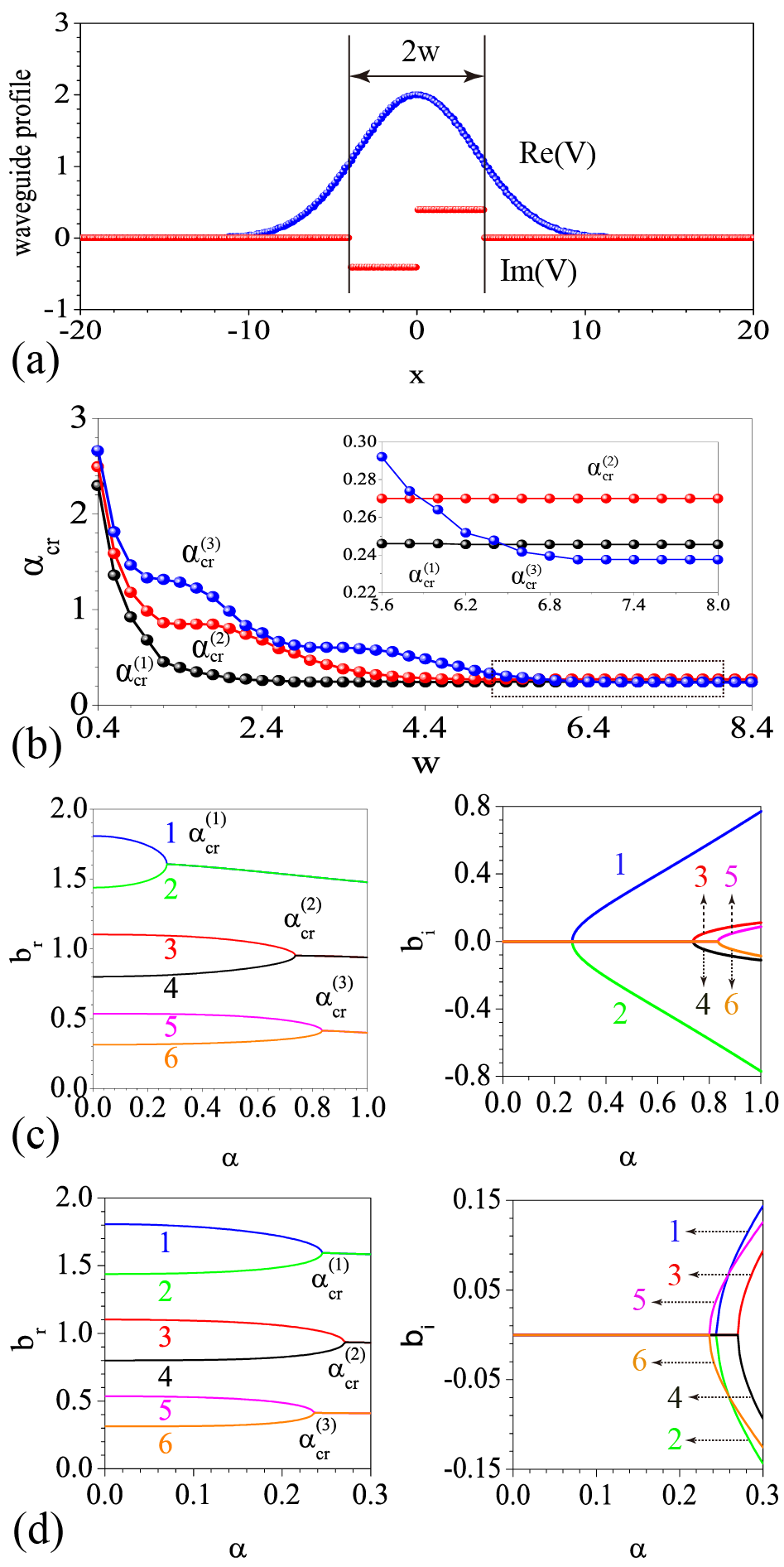}}\vspace{-0.025\textwidth}\\
\caption{\label{fig:Fig2} (Color online) (a) Real(Re(V)) and imaginary (Im(V)) part of the waveguide profile,
with $\text{Re(V)}= p \exp(x^2/d^2)$ for all x, while $\text{Im(V)}=ix/|x|\alpha$ for $x\in (-w,w)$, and 0 for otherwise. (b) Dependence of $\alpha_{cr}$ of three mode pairs on $w$. The inset is the enlargement of the dashed-box portion.  The dependence of mode propagation constants on $\alpha$ for $w=2.2$ and $w=7.4$ is shown in (c) and (d), respectively. In all the cases, $p=2, d=5.$}
\end{figure}

\section{$\mathcal{PT}$-SYMMETRY BREAKING POINT OF DIFFERENT MODE PAIRS}
Finally we shall compare the $\mathcal{PT}$-symmetry breaking point for different mode pairs. In canonical waveguiding geometries such as the Gaussian waveguides
considered above, one finds that, with the increase of gain/loss coefficient, the lowest-order mode pair first breaks symmetry, then higher-order pairs break
theirs successively. This property is clearly seen in Fig.~2 and Fig.~5, as $\alpha_{\text{cr}}^{(1)}<\alpha_{\text{cr}}^{(2)}<\alpha_{\text{cr}}^{(3)}<...$. Thus, for
 some specific gain and loss level, while the lower-order mode pairs are already symmetry broken, the higher-order pairs might still maintain their symmetry.
  Figure 6 shows the propagation simulation for fundamental/dipole [Fig.~6(a,b)] and tripole/quadrupole [Fig.~6(c,d)] pairs. Note that a same $\alpha$ value
  ($\alpha=0.175$) is used in Fig.~6(b) and (c). However, as $\alpha_{\text{cr}}^{(1)}<0.175<\alpha_{\text{cr}}^{(2)}$, the modes bifurcating from
  fundamental/dipole mode pair experience either amplification[Fig.~6(b)] or decay during propagation, while the modes in the tripole/quadrupole pair still propagate
  in a stationary fashion[Fig.~6(c)]. We note that, the fact that the symmetry of the tripole is maintained while that of the two lower-order modes is already
   broken was indicated in \cite{new1}.
%\begin{figure}[htbp]
%\centering
%\subfigure{\label{fig:fig7a}\includegraphics[width=0.44\textwidth]{fig7a.png}}\vspace{-0.025\textwidth}\\
%\subfigure{\label{fig:fig6a}\includegraphics[width=0.24\textwidth]{fig6a.png}}\hspace{-0.01\textwidth}
%\subfigure{\label{fig:fig6b}\includegraphics[width=0.24\textwidth]{fig6b.png}}\vspace{-0.025\textwidth}\\
%\subfigure{\label{fig:fig6a}\includegraphics[width=0.24\textwidth]{fig7b.png}}\hspace{-0.01\textwidth}
%\subfigure{\label{fig:fig6b}\includegraphics[width=0.24\textwidth]{fig7b.png}}
%\caption{\label{fig:Fig2} (Color online) Dependence of mode propagation constants on $\alpha$ for a waveguide given by  $V(x)=p\exp(x^2/d^2)+ix/|x|\alpha,p=2,d=5$. The inset in (b) shows the real(Re(V)) and imaginary(Im(V)) part of the waveguide profile.}
%\end{figure}

One might explain that the postpone in the symmetry breaking of the higher-order mode pair is due to the fact that higher-order modes are more spatially extended,
 and thus the \emph{effective} gain and loss strength they feel are weaker than the lower-order pairs do, therefore a larger gain/loss level is required to drive
  higher-order pairs into symmetry-breaking phases(the effective gain/loss strength is given by the spatially weighted average of the imaginary part of the complex
   potential over the modal field profile). However, we find that this argument is not always true, and the higher-order mode pairs may also break
   symmetry \emph{earlier} than lower-order pairs do. A profound example is shown in Fig.~7, where the real part of the potential is still a Gaussian one,
   while the gain/loss modulation is a step function defined within a finite region with width being $2w$ [Fig.~7(a)]. We examine the dependence of symmetry-breaking
    points for different mode pairs of the structure on the width of gain/loss region, $w$, and the result is shown Fig.~7(b). It shows that, when the gain
     and loss region is narrow(compared with the width of the mode profiles), the breaking points of all mode pairs are nearly the same; with the increasing $w$,
     it becomes evident that higher-order mode pairs break symmetry later than lower mode pairs do(see, for example, Fig.~7(c) for $w=2.2$). Interestingly,
     when $w$ is increased further($w>6.4$), the third mode pair is found to firstly break symmetry, followed by the first mode pair, and then the second mode
     pair(Fig.~7(d)). This situation remains true even when $w\rightarrow \infty$. Clearly, for such a very spatially extended gain/loss region, the effective gain
      and loss strengths for all mode pairs are the same, and still, different mode pairs break symmetry at different point. Finally, we should mention that,
      the observed property£¨the mode pairs of higher orders could break symmetry earlier than those of lower orders do), is not caused by the the piecewise nature of
      the imaginary potential considered in Figure 7, and a similar picture is observed from our simulations for other smoothly varying potentials, too.
\section {CONCLUSION}
In conclusion, we have put forward a systematic study on the properties of $\mathcal{PT}$ symmetry for multimode waveguides. We have
revealed that waveguide modes with successive orders 2\emph{m}-1 and 2\emph{m} form a mode pair as they gradually evolve into the same mode
 and experience symmetry breaking simultaneously. For waveguides that support an odd number of guided modes, the increase of gain and loss
  coefficient gives birth to a new higher-order mode which pairs with the already existing highest-order mode, and then go to a symmetry breaking
  together. Depending on the specific realizations of $\mathcal{PT}$-symmetric potentials, the breaking point of the higher-order mode pair can be
  later or earlier than those of the lower-order pairs.

\begin{acknowledgments}
The authors thank B. Malomed and Y. Kartashov for useful discussions. F.Y acknowledges support from the National Natural Scientific Funding of China(No. 11104181).
\end{acknowledgments}


%{osajnl}

\end{document}